\def\Journal#1#2#3#4{{#1} {\bf #2}, #3 (#4)}

\def\LNC{\em Lett.\ Nuovo Cimento}

\def\NPB{{\em Nucl.\ Phys.} B}
\def\PLA{{\em Phys.\ Lett.}  A}
\def\PLB{{\em Phys.\ Lett.}  B}
\def\PRL{\em Phys.\ Rev.\ Lett.}
\def\PR{\em Phys.\ Rev.}
\def\PRA{{\em Phys.\ Rev.} A}
\def\PRB{{\em Phys.\ Rev.} B}

\def\PRE{{\em Phys.\ Rev.} E}
\def\RMP{\em Rev.\ Mod.\ Phys.}

\def\JMP{\em J.\ Math.\ Phys.}
\def\JCP{\em J.\ Chem.\ Phys.}
\def\JCompP{\em J.\ Comp.\ Phys.}
\def\CPL{\em Chem.\ Phys.\ Lett.}
\def\JASA{\em J.\ Am.\ Stat.\ Ass.}
\def\SR{\em SIAM Review}
\def\JPSJ{\em J.\ Phys.\ Soc.\ Jpn.}
\def\AP{\em Ann.\ Phys.}
\def\SPU{\em Sov.\ Phys.\ Usp.}
\def\AJP{\em Am.\ J.\ Phys.}

\documentclass{article}

\usepackage{amsmath}
\begin{document}

\title{The Feynman Path Goes Monte Carlo} 

\author{Tilman Sauer\thanks{To appear in: \emph{Fluctuating Paths and
      Fields. Festschrift Dedicated to Hagen Kleinert}, Eds. W. Janke,
    A. Pelster, H.-J. Schmidt, and M. Bachmann
    (World Scientific, Singapore, 2001).}\\
  {\small History and Philosophy of Science}\\
  {\small University of Berne, Sidlerstrasse 5, CH-3012 Berne, Switzerland}\\
  {\small E-mail: tilman.sauer@philo.unibe.ch}}

\maketitle

\begin{abstract}
  Path integral Monte Carlo (PIMC) simulations have become an
  important tool for the investigation of the statistical mechanics of
  quantum systems. I discuss some of the history of applying the Monte
  Carlo method to non-relativistic quantum systems in path-integral
  representation. The principle feasibility of the method was well
  established by the early eighties, a number of algorithmic
  improvements have been introduced in the last two decades.
\end{abstract}

\section{Introduction}

Feynman's\index{FEYNMAN, R.P.} classic paper of 1948 presented a {\em
  Space-Time Approach to Non-Relativistic Quantum
  Mechanics}, \cite{sauer:fey48} or, in Hagen
Kleinert's\index{KLEINERT, H.}  words, ``an all-time global approach
to the calculation of quantum mechanical amplitudes.''  Within the
philosophy of this approach, we must find, as Kleinert often stressed,
``all properties, including the Schr\"odinger wave functions, from the
globally determined time displacement amplitude.'' \cite{sauer:kle90}
The Feynman path, governed by the classical Lagrangian of the quantum
system, is the very object we need to study if we want to establish a
truly independent alternative to Schr\"odinger's equation.  In
avoiding the operator formalism, the sum over paths provides an
independent conceptual link between quantum and classical mechanics.

Another attractive feature of the path integral formulation of quantum
mechanics is the bridge it allows to establish between quantum
mechanics and statistical mechanics. Technically, the oscillating
exponential of the time-displace\-ment amplitude turns into a positive
Boltzmann weight if the paths are expressed in imaginary time.
The quantum mechanical propagator thus turns into a quantum
statistical density matrix. It is this very feature which allows the
application of methods of classical statistical mechanics, notably the Monte
Carlo method, to quantum systems.

The Monte Carlo method\index{Monte Carlo method} came into being
roughly around the same time as the Feynman path. Anecdotally, the
idea of gaining insight into a complex phenomenon by making various
trials and studying the proportions of the respective outcomes
occurred to Stanislaw Ulam\index{ULAM, S.} while playing solitaire
during an illness in 1946 \cite{sauer:asp89}. The immediate application
was, of course, the problem of neutron diffusion studied in Los Alamos
at that time. The name of the procedure first appeared in print in a
classic paper by Metropolis\index{METROPOLIS, N.} and Ulam\index{ULAM,
  S.} in 1949 \cite{sauer:mu49}, where the authors explicitly mentioned
that the method they presented as a statistical approach to the study
of integro-differential equations would sometimes be referred to as
the Monte Carlo method. In classical statistical mechanics it quickly
became a standard calculational tool.

\section{Where's Monte Carlo?}

The object of interest in Monte Carlo evaluations of Feynman's path
integral is the quantum statistical partition function ${\cal Z}$,
given, in operator language, as the trace of the density operator
$\exp(-\beta \hat{H})$ of the canonical ensemble ($\beta=1/k_{\rm
  B}T$) associated with a Hamilton operator describing $N$ particles
of mass $m_i$ moving under the influence of a potential $V$,
\begin{equation}
\hat{H} = \sum_{i=1}^N \frac{\hat{\vec{p}_i}^2}{2m_i} 
                    + V(\hat{\vec{r}}_1, \dots, \hat{\vec{r}}_N).
\end{equation}
Expressed as a Feynman integral, the density matrix elements read
\begin{equation}
\langle{\bf r}|\exp(-\beta \hat{H})|{\bf r'}\rangle
= \int\limits_{{\bf r}(0)={\bf r}}^{{\bf r}(\hbar\beta)
={\bf r'}} {\cal D}{\bf r}(\tau)\exp
 \left\{-\frac{1}{\hbar}\int\limits_0^{\hbar\beta} L
\left(\{\dot{\vec{r}}_i(\tau), \vec{r}_i(\tau)\}\right) d\tau\right\}
\label{sauer:eq:density_matrix}
\end{equation}
where ${\bf r}\equiv \{\vec{r}_1, \dots, \vec{r}_N\}$, $L$ denotes
the classical Lagrangian
\begin{equation}
L\left(\{\dot{\vec{r}}_i(\tau), \vec{r}_i(\tau)\}\right) =
\sum_{i=1}^ N 
     \frac{m_i}{2}\dot{r}_i^2 + V(\vec{r_1}, \dots, \vec{r}_N(\tau))
\end{equation}
expressed in imaginary time $\tau$.%
\footnote{There have been attempts to apply the Monte Carlo method to
  path integrals also for real time. However, due to the oscillating
  exponential one then has to deal with problems of numerical
  cancellation, and it is much harder to obtain results of some
  numerical accuracy. Therefore, I shall here restrict myself to Monte
  Carlo work in imaginary time.}  The particles are assumed to be
distinguishable. To evaluate the trace, we only need to set ${\bf
  r}={\bf r'}$ and integrate over ${\bf r}$. To take into account Bose
or Fermi statistics for indistinguishable particles, the partition
function splits into a sum of the direct Boltzmann part and parts with
permuted endpoints.

The right hand side of Eq.~(\ref{sauer:eq:density_matrix}) is a path
integral for the $3N$ functions ${\bf r}$. The idea of a Monte Carlo
evaluation of this quantity is to sample these paths stochastically
and to get (approximate) information about the quantum statistics of
the system by averaging over the finite set of paths generated in the
sampling process.

Monte Carlo data always come with error bars and, in general, the
errors associated with numerical Monte Carlo data stem from two
distinct sources. A \emph{systematic} error of Monte Carlo
evaluations of the path integral follows from the need to identify the
paths by a finite amount of computer information.  This can be done by
discretizing the paths at some set of points in the interval
$(0,\hbar\beta)$. For a single particle moving in one dimension, the
simplest discrete time approximation for $L$ time slices reads
($\epsilon=\hbar\beta/L$)
\begin{eqnarray}
&\langle x|&\exp(-\beta \hat{H})|x'\rangle = \notag\\
&& \lim_{L\rightarrow\infty} \frac{1}{A}\prod_{j=1}^{L-1}
\left[\int\frac{dx_j}{A}\right]\exp\left\{-\frac{1}{\hbar}\sum_{j=1}^L
\left[\frac{m}{2}\frac{(x_j-x_{j-1})^2}{\epsilon}
+\epsilon V(x_{j-1})\right]\right\}
\label{sauer:eq:dta}
\end{eqnarray}
where $A=(2\pi\hbar\epsilon/m)^{1/2}$ and $x_0=x$ and $x_L=x'$.
Alternatively, one may expand the individual paths in terms of an
orthogonal function basis, e.g.\ by the Fourier decomposition,
\begin{equation}
x(\tau) = x+\frac{(x'-x)\tau}{\hbar\beta} 
+ \sum_{k=1}^{\infty}a_{k}\sin\frac{k\pi \tau}{\hbar\beta},
\end{equation}
and express the density matrix as
\begin{eqnarray}
\langle x|\exp(-\beta \hat{H})|x'\rangle &=& \lim_{L'\rightarrow\infty}
J\exp\left\{-\frac{m}{2\hbar^2\beta}(x-x')^2\right\}\times\notag\\
\times\int\prod_{k=1}^{L'}&& da_k\exp\left\{-\frac{a_k^2}{2\sigma_k^2}\right\}
\times
\exp\left\{-\frac{1}{\hbar}\int\limits_0^{\hbar\beta} V(x(\tau))d\tau\right\}
\end{eqnarray}
where $\sigma_k=[2\hbar^2\beta/m(\pi k)^2]^{1/2}$ and $J$ is the
Jacobian of the transformation from the integral over all paths to the
integral over all Fourier coefficients. A systematic error then arises
from the loss of information by the finite number $L$ of points
$x_i$ on the discretized time axis or by the finite number $L'$ of
Fourier components $a_k$ that are taken into account in the Monte
Carlo sampling of the paths.

The other error source of Monte Carlo data is the \emph{statistical}
error due to the finite number $N_m$ of paths that form the sample
used for evaluating the statistical averages. To make matters worse,
the probability of configurations is, in general highly peaked, making
an independent sampling of paths highly inefficient in most cases. The
remedy is to introduce some way of ``importance sampling'' where
configurations are generated according to their probability given by
the exponential in Eq.~(\ref{sauer:eq:density_matrix}). Statistical
averages may then be computed as simple arithmetic means. A way to
achieve this is by constructing Markov chains where transition
probabilities between configuration are constructed that allow to
generate a new configuration from a given one such that in the limit
of infinitely many configurations the correct probability distribution
of paths results. A very simple and universally applicable algorithm
to set up such a Markov chain is the Metropolis
algorithm\index{Metropolis algorithm} introduced in
1953 \cite{sauer:mrrtt53}. Here a new configuration is obtained by
looking at some configuration with only one variable changed and
accepting or rejecting it for the sample on the basis of a simple rule
that depends only on the respective energies of the two
configurations. The advantages of importance sampling on the basis of
Markov chains are obtained on the cost that, in general, successive
configurations are not statistically independent but autocorrelated.
The crucial quantity is the integrated autocorrelation time
$\tau_{{\cal O}}^{\rm int}$ of a quantity of interest ${\cal
  O}=\langle\overline{\cal O}\rangle$ with $\overline{\cal O} =
(1/N_m)\sum_{i=1}^{N_m}{\cal O}_i$ and ${\cal O}_i$ computed for each
path $i$ in the sample.  It enters the statistical error estimate
$\Delta_{\cal O}$ for expectation values of ${\cal O}$ computed from a
Monte Carlo sample of $N_m$ autocorrelated configurations as
\begin{equation}
\Delta{\overline{\cal O}}=\sqrt{\frac{\sigma_{\cal O_i}^2}{N_m}}
\sqrt{2\tau_{\cal O}^{\rm int}}
\end{equation}
where $\sigma_{{\cal O}_i}^2$ is the variance of ${\cal O}_i$.

With Monte Carlo generated samples of Feynman paths one can thus
``measure'' thermodynamic properties of quantum systems like the
internal energy and the specific heat, but also gain more detailed
information about correlation functions, probability distributions and
the like. In the low-temperature limit, $\beta\rightarrow\infty$,
quantum mechanical ground state properties are recovered.

\section{Blazing Trails}

A pioneer in the application of the Monte Carlo method to physics
problems, notably by applying it to the Ising model, was Lloyd D.\ 
Fosdick. He appears to have also been one of the first to
consider the stochastic sampling of paths.  In 1962, he considered the
possibility of sampling paths \cite{sauer:fos62} for what he called the
conditional Wiener integral, i.e.\ the Wiener integral for fixed end
points.  As a toy example he investigated the expectation value of the
functional $\exp\left[-\int_0^1\int_0^1\tau\tau'x(\tau)x(\tau')d\tau
  d\tau'\right]$ for a conditional Wiener process, and, more
generally, for the quantity $\exp\left[-\int_0^{\beta}Vd\tau\right]$,
i.e.\ he considered direct computation of the partition function for a
quantum particle moving in a potential $V$. He introduced a Fourier
decomposition of the paths and generated these by direct Monte Carlo
sampling of the Fourier components as Gaussian stochastic variables.
He did some explicit sampling of his toy model to demonstrate the
feasibility of the method but a theoretical consideration of the
one-dimensional harmonic oscillator was not considered worthwhile to
be put on the computer, even though Fosdick at the time was at the
University of Illinois and had access to the university's ILLIAC
computer. His examples were primarily used to investigate the
principle feasibility and possible accuracy obtainable by the method.
Instead, Fosdick went along to consider a pair of two identical
particles and presented some numerical results for this problem.
Continuation of the work on the two-particle problem together with a
graduate student led to the publication of a paper on the Slater sum,
i.e.\ the diagonal density matrix elements, for He$^4$ in
1966 \cite{sauer:fj66}, and on three-particle effects in the pair
distribution function for He$^4$ in 1968 \cite{sauer:jf68}. In the same
year \cite{sauer:fos68}, Fosdick elaborated on the numerical method in a
report on the Monte Carlo method in quantum statistics in the SIAM
review.  Instead of sampling the Fourier components he now used
the discrete time approximation of the paths. Sampling of $x_i$ at the
discrete points was done using a trick that later would gain
prominence in PIMC simulations in an algorithm called
staging.\index{Staging algorithm} The idea is to express the
discretized kinetic term in the relative probablity density
$p(x_i|x_{i-1},x_{i+1}) = (1/2\pi\epsilon)\exp\left[-(x_i-x_{i-1})^2/2
  \epsilon-(x_{i+1}-x_i)^2/2\epsilon\right]$ as
$-(x_i-\bar{x}_i)^2/2\sigma^2+(x_{i-1}-x_{i+1})^2/4\epsilon$ with
$\bar{x_i}=(x_{i-1}+x_{i+1})/2$ and to sample $(x_i-\bar{x_i})/\sigma$
as an independent Gaussian variable. The procedure could be iterated
recursively for all $x_{i}$ and thus allowed to obtain statistically
independent paths which were used to ``measure'' the potential energy
term $\exp\left[-\int_0^{\beta}V(x(\tau))d\tau\right]$.

In 1969, Lawande\index{LAWANDE, S.V.}, Jensen\index{JENSEN, C.A.}, and
Sahlin\index{SAHLIN, H.L.} introduced Metropolis sampling of the paths
in discrete time, broken line approximation \cite{sauer:ljs69a}. They
investigated the ground state wave functions of simple one-dimensional
problems (harmonic oscillator, square well, and Morse potential) and,
theoretically, also addressed the problem of extracting information
about excited energies and of simulating many particle problems. In a
follow-up paper \cite{sauer:ljs69b} they presented investigations of
the Coulomb problem\index{Coulomb problem} using Monte Carlo
simulations of the path integral in polar coordinates. Not
surprisingly, the singularity at the origin had to be avoided by
artificial constraints and the authors admitted that a more rigorous
justification of their procedure was called for. The path integral was
later solved exactly by Duru and Kleinert in 1979 \cite{sauer:dk79}. It
became clear that there were fundamental problems with such
singularities in time-sliced path integrals \cite{sauer:kle90}.

Little activity is recorded in the seventies, and I am only aware of a
brief theoretical consideration of the possibility of Monte Carlo
sampling of paths in a paper by Morita\index{MORITA, T.} on the
solution of the Bloch equation for many particle systems in terms of
path integrals from 1973 \cite{sauer:mor73}. The paper is cited in a later
one by J.A.~Barker\index{BARKER, J.A.} published in 1979 \cite{sauer:bar79}
in which the one-dimensional particle in a box\index{Particle in a
  box} is considered, and numerical estimations of the ground state
energy and wave function are presented.  The data were obtained by
introducing image sources to take account of the boundary conditions
of the box and using Metropolis sampling of the broken line
approximation of the paths.  Incidentally, the analytic solution of
this problem, i.e.\ of the path integral for the particle in the box
was given by Janke\index{JANKE, W.} and Kleinert\index{KLEINERT, H.} 
almost simultaneously \cite{sauer:jk79}.
Barker also computed distribution functions for the problem of two
particles in a box.

Very much in the spirit of Lawande et al.\ but possibly unaware of
this work, Creutz\index{CREUTZ, M.} and Freedman\index{FREEDMAN, B.}
published a didactic paper on {\em a statistical approach to quantum
  mechanics} in 1981 \cite{sauer:cf81}. They, too, performed Metropolis
sampling of paths in the broken line approximation and studied the
energies and ground state wave functions of the one-dimensional
harmonic oscillator. The background of these authors were Monte Carlo
studies of gauge field theories, and the paper was meant as an attempt
to better understand the Monte Carlo method by applying it to simple
one-degree-of-freedom Schr\"odinger problems. It still is a useful
introduction to the basics of the technique, and in particular it
presents a brief primer on the theory of Markov chains underlying the
Metropolis algorithm. To compute energies they introduced an
alternative estimator by invoking the virial theorem. They also
studied double well problems, presenting snap shot pictures of
double-kink instanton paths.\index{Instantons} The problem of
determining the energy level splitting was addressed by computing
correlation functions.

The papers by Lawande et al.\ and by Creutz and Freedman appear to
have been cited very rarely, possibly because they presented their
work as being only of pedagogic value and not so much because the
Monte Carlo method could be a useful method to obtain numerical
results for Schr\"odinger problems which, in real life, should be
handled by numerical methods more suitable in this simple case. These
remarks also hold for work published a little later by
Shuryak\index{SHURYAK, E.V.} \cite{sauer:sz84,sauer:shu84}.

Fosdick's\index{FOSDICK, L.D.} work from 1962 was done very much at
the forefront of the technological possibilities of high-speed
computing at the time. By the mid-eighties, path integral simulations
of simple quantum mechanical problems had become both conceptually and
technically ``easy.'' Indeed, the exposition by Creutz and Freedman
was already written in an introductory, didactic manner, and in 1985
the simulation of the one-particle harmonic oscillator was explicitly
proposed as an undergraduate project, to be handled on a Commodore
CBM3032 microcomputer, in a paper published in the American Journal of
Physics \cite{sauer:mac85}.

\section{Speeding up}

The feasibility of evaluating the quantum statistical partition
function of many-particle systems by Monte Carlo sampling of paths was
well established by the early eighties and the method began to be
applied to concrete problems, in particular in the chemical physics
literature. It had also become clear that the method had severe
restrictions if numerical accuracy was called for. In addition to the
statistical error inherent to the Monte Carlo method, a systematic
error was unavoidably introduced by the necessary discretization of
the paths. Attempts to improve the accuracy by algorithmic
improvements to reduce both the systematic and the statistical errors
were reported in subsequent years. The literature is abundant and
rather than trying to review the field I shall only indicate some
pertinent paths of development.

In Fourier PIMC methods,\index{Fourier path integral Monte Carlo}
introduced in 1983 in the chemical physics context by Doll\index{DOLL,
  J.D.} and Freeman\index{FREEMAN, D.L.} \cite{sauer:df84,sauer:fd84}, the
systematic error arises from the fact that only a finite number of
Fourier components are taken into account. Here the systematic error
could be reduced by the method of partial averaging \cite{sauer:dcf85,sauer:cfd86}.

In discrete time approximations arising from the short-time propagator
or, equivalently, the high-temperature Green's function various
attempts have been made to find more rapidly converging formulations.
Among these are attempts to include higher terms in an expansion of
the Wigner-Kirkwood\index{Wigner-Kirkwood expansions} type, i.e.\ an
expansion in terms of $\hbar^2/2m$.  Taking into account the first
term of such an expansion would imply to replace the potential term
$\epsilon V(x_{j-1})$ in (\ref{sauer:eq:dta})
by \cite{sauer:mm88,sauer:mm89,sauer:bgm90}
\begin{equation}
\epsilon V(x_{j-1}) \rightarrow \frac{\epsilon}{x-x'}\int_{x}^{x'}dyV(y).
\end{equation}
This improves the convergence of the density matrix
(\ref{sauer:eq:dta}) (from even less than ${\cal
  O}(1/L)$) \cite{sauer:mm88} to ${\cal O}(1/L^2)$.  For the full
partition function, the convergence of the simple discretization
scheme is already of order ${\cal O}(1/L^2)$ since due to the cyclic
property of the trace, the discretization $\epsilon V(x_{j-1})$ is
then equivalent to a symmetrized potential term $\epsilon
(V(x_{j-1})+V(x_j))/2$. The convergence behaviour of these
formulations follows from the Trotter decomposition
formula,\index{Trotter formula}
\begin{equation}
e^{-(A+B)}=\left[e^{-\frac{A}{L}}e^{-\frac{B}{L}}\right]^L 
+ {\cal O}(\frac{1}{L})
=\left[e^{-\frac{A}{2L}} e^{-\frac{B}{L}}e^{-\frac{A}{2L}}\right]^L 
+ {\cal O}(\frac{1}{L^2}),
\end{equation}
valid for non-commuting operators $A$ and $B$ in a Banach
space \cite{sauer:suz85}, identifying $A$ with the kinetic energy
$\beta\sum\hat{\vec{p}}_i^2/2m_i$ and $B$ with the potential energy
$\beta V(\{\hat{x}_i\})$. More rapidly converging discretization
schemes were investigated on the basis of higher-order decompositions.
Unfortunately, a direct, ``fractal'' decomposition \cite{sauer:suz90} of the
form
\begin{equation}
e^{-(A+B)}=\lim_{L\rightarrow\infty}
\left[
e^{\alpha_1 \frac{A}{L}}
e^{\beta_1 \frac{B}{L}}  
e^{\alpha_2\frac{A}{L}}
e^{\beta_2\frac{B}{L}}
\dots
\right]^L, \,\,\, 
\sum\alpha_i=\sum\beta_i=1
\end{equation}
inevitably leads to negative coefficients
for higher decompositions \cite{sauer:suz91} and is therefore not amenable to
Monte Carlo sampling of paths \cite{sauer:js92a}. Higher-order Trotter
decomposition schemes involving commutators have proven to be more
successful \cite{sauer:dd83,sauer:ti84,sauer:lb87,sauer:ktl88}. In particular, 
a decomposition of the form 
\begin{equation}
{\cal Z} =  \lim_{L\rightarrow\infty}{\rm Tr} 
\left[e^{-\frac{A}{2L}}e^{-\frac{B}{2L}} e^{-\frac{[[B,A],B]}{24L^3}}
  e^{-\frac{B}{2L}}e^{-\frac{A}{2L}}\right]^L, 
\end{equation}
derivable by making use of the cyclic
property of the trace, is convergent of order ${\cal O}(1/L^4)$ and
amounts to simply replacing the potential $\epsilon V$ in
(\ref{sauer:eq:dta}) by an effective potential \cite{sauer:ti84} 
\begin{equation}
V_{\rm eff} = V + \frac{(\beta\hbar)^2}{24mL^2}(V')^2.
\end{equation}

Another problem for the numerical accuracy of PIMC simulations arises
from the analog of the ``critical slowing down''
\index{Critical slowing down} problem well-known for local update algorithms at
second-order phase transitions in the simulation of spin systems and
lattice field theory. Since the correlations $\langle
x_jx_{j+k}\rangle$ between variables $x_j$ and $x_{j+k}$ in the
discrete time approximation only depend on the temperature and on the
gaps between the energy levels and not, or at least not appreciably,
on the discretization parameter $\epsilon$, the correlation length
$\zeta$ along the discretized time axis always diverges linearly with
$L$ when measured in units of the lattice spacing $\epsilon$. Hence in
the continuum limit of $\epsilon\rightarrow 0$ with $\beta$ fixed or,
equivalently, of $L\rightarrow\infty$ for local, importance sampling
update algorithms, like the standard Metropolis algorithm, a slowing
down occurs because paths generated in the Monte Carlo process become
highly correlated.  Since for simulations using the Metropolis
algorithm autocorrelation times diverge as \cite{sauer:js92b} $\tau_{\cal
  O}^{\rm int}\propto L^z$ with $z\approx 2$ the computational effort
(CPU time) to achieve comparable numerical accuracy in the continuum
limit $L\rightarrow \infty$ diverges as $L\times L^z=L^{z+1}$.

To overcome this drawback, ad hoc algorithmic modifications like
introducing collective moves of the path as a whole between local
Metropolis updates were introduced then and again. One of the earliest
more systematic and successful attempts to reduce autocorrelations
between successive path configurations was introduced in 1984 by
Pollock\index{POLLOCK, E.L.} and Ceperly \cite{sauer:pc84}.
\index{CEPERLEY D.M.} Rewriting the discretized path integral, their method
essentially amounts to a recursive transformation of the variables
$x_i$ in such a way that the kinetic part of the energy can be taken
care of by sampling direct Gaussian random variables and a Metropolis
choice is made for the potential part. The recursive transformation
can be done between some fixed points of the discretized paths, and
the method has been applied in such a way that successively finer
discretizations of the path were introduced between neighbouring
points. Invoking the polymer analog of the discretized path this
method was christened the ``staging'' algorithm\index{Staging algorithm} 
by Sprik,\index{SPRIK, M.} Klein,\index{KLEIN, M.L.} and
Chandler\index{CHANDLER, D.} in 1985 \cite{sauer:skc85}.

The staging algorithm decorrelates successive paths very effectively
because the whole staging section of the path is essentially sampled
independently. In 1993, another explicitly non-local update was
applied to PIMC simulations \cite{sauer:js92b,sauer:js93} by transferring the
so-called multigrid method\index{Multigrid method} known from the
simulation of spin systems.  Originating in the theory of numerical
solutions of partial differential equations, the idea of the multigrid
method is to introduce a hierarchy of successively coarser grids in
order to take into account long wavelength fluctuations more
effectively. Moving variables of the coarser grids then amounts to a
collective move of neighbouring variables of the finer grids, and the
formulation allows to give a recursive description of how to cycle
most effectively through the various levels of the multigrid.
Particularly successful is the so-called W-cycle. Both the staging
algorithm and the multigrid W-cycle have been shown to beat the
slowing down problem in the continuum limit completely by reducing the
exponent $z$ to $z\approx 0$ \cite{sauer:js96}.

Another cause of severe correlations between paths arises if the
probability density of configurations is sharply peaked with high
maxima separated by regions of very low probability density. In the
statistical mechanics of spin systems this is the case at a
first-order phase transition. In PIMC simulations the problem arises
for tunneling situations like, e.g., for a double well potential with
a high potential barrier between the two wells. In these cases, an
unbiased probing of the configuration space becomes difficult because
the system tends to get stuck around one of the probability maxima. A
remedy to this problem is to simulate an auxiliary distribution that
is flat between the maxima and to recover the correct Boltzmann
distribution by an appropriate reweighting of the sample. The
procedure is known under the name of umbrella sampling\index{Umbrella
  sampling} or multicanonical sampling.\index{Multicanonical sampling}
It was shown to reduce autocorrelations for PIMC simulations of a
single particle in a one-dimensional double well, and it can also be
combined with multigrid acceleration \cite{sauer:js94}.

The statistical error associated with a Monte Carlo estimate of an
observable ${\cal O}$ cannot only be reduced by reducing
autocorrelation times $\tau^{\rm int}_{\cal O}$. If the observable can
be measured with two different estimators $U_i$ that yield the same
mean $U_i^{(L)}=\langle U_i\rangle$ with ${\cal O} =
\lim_{L\rightarrow\infty}U_i^{(L)}$, the estimator with the smaller
variance $\sigma^{2}_{U_i}$ is to be preferred. Straighforward
differentiation of the discretized path integral (\ref{sauer:eq:dta})
leads to an estimator of the energy\index{Energy estimation} that
explicitly measures the kinetic and potential  parts of the energy by
\begin{equation}
U_{\rm k} = \frac{L}{2\beta}-\frac{m}{2L}
\sum\left(\frac{x_j-x_{j-1}}{\epsilon}\right)^2 
+ \frac{1}{L}\sum_{i=1}^LV(x_i).
\end{equation}
The variance of this so-called ``kinetic'' energy estimator diverges
with $L$. Another estimator can be derived by invoking the path analog
of the virial theorem
\begin{equation}
\frac{L}{2\beta}-\frac{m}{2}\left
\langle\left(\frac{x_j-x_{j-1}}{\epsilon}\right)^2\right\rangle
=\frac{1}{2}\langle x_jV'(x_j)\rangle, 
\end{equation}
and the variance of the ``virial'' estimator 
\begin{equation}
U_{\rm v} = \frac{1}{2L}\sum_{i=1}^Lx_iV'(x_i) 
+ \frac{1}{L}\sum_{i=1}^L V(x_i)
\end{equation}
does not depend on $L$. In the early eighties, investigations of the
``kinetic'' and the ``virial'' estimators focussed on their
variances \cite{sauer:hbb82,sauer:pr84,sauer:ti84}. Some years later,
it was pointed out \cite{sauer:gj88} that a correct assessment of the
accuracy also has to take into account the autocorrelations, and it
was demonstrated that for a standard Metropolis simulation of the
harmonic oscillator the allegedly less successful ``kinetic''
estimator gave smaller errors than the ``virial'' estimator. In 1989
it was shown \cite{sauer:cb89} that conclusions about the accuracy also
depend on the particular Monte Carlo update algorithm at hand since
modifications of the update scheme such as inclusion of collective
moves of the whole path affect the autocorrelations of the two
estimators in a different way. A careful comparison of the two
estimators which disentangles the various factors involved was given
only quite recently \cite{sauer:js97}. Here it was also shown that a
further reduction of the error may be achieved by a proper combination
of both estimators without extra cost.

\section{Concluding Remarks}

Application of the Monte Carlo method to quantum systems is not
restricted to direct sampling of Feynman paths but this method has
attractive features. It is not only conceptually suggestive but also
allows for algorithmic improvements that help to make the method
useful even when the problems at hand requires considerable numerical
accuracy.  However, algorithmic improvements like the ones alluded to
above have tended to be proposed and tested mainly for simple,
one-particle systems. On the other hand, the power of the Monte Carlo
method is, of course, most welcome in those cases where analytical
methods fail. For more complicated systems, however, evaluation of the
algorithms and control of numerical accuracy is also more difficult.
Only recently, a comparison of the efficiency of Fourier- and discrete
time-path integral Monte Carlo for a cluster of 22 hydrogen molecules
was presented \cite{sauer:cgc98}---and debated \cite{sauer:df99,sauer:cgc99}.
Nevertheless, path integral Monte Carlo simulations have become an
essential tool for the treatment of strongly interacting quantum
systems, like, e.g., the theory of condensed helium \cite{sauer:cep95}.

\section*{Acknowledgments}

I wish to thank Wolfhard Janke for instructive and enjoyable
collaboration.

%\printindex

\end{document}